\documentclass{emulateapj}
\usepackage{lscape,color}
\shorttitle{Far-IR spectra of compact sources in the SMC}
\shortauthors{van Loon et al.}
\begin{document}
\title{A {\it Spitzer Space Telescope} far-infrared spectral atlas of compact
sources in the Magellanic Clouds. II. The Small Magellanic Cloud}
\author{Jacco Th. van Loon\altaffilmark{1}}
\email{jacco@astro.keele.ac.uk}
\author{Joana M. Oliveira\altaffilmark{1}}
\author{Karl D. Gordon\altaffilmark{2}}
\author{G. C. Sloan\altaffilmark{3}}
\author{C. W. Engelbracht\altaffilmark{4}}
\affil{
$^{1}$ Astrophysics Group, Lennard-Jones Laboratories, Keele University,
Staffordshire ST5 5BG, UK\\
$^{2}$ Space Telescope Science Institute, 3700 San Martin Drive, Baltimore, MD
21218, USA\\
$^{3}$ Department of Astronomy, Cornell University, Ithaca, NY 14853, USA\\
$^{4}$ Steward Observatory, University of Arizona, 933 North Cherry Avenue,
Tucson, AZ 85721, USA
}
\begin{abstract}
We present far-infrared spectra, $\lambda$=52--93 $\mu$m, obtained with the
{\it Spitzer Space Telescope} in the Spectral Energy Distribution mode of its
MIPS instrument, of a selection of luminous compact far-infrared sources in
the Small Magellanic Cloud. These comprise nine Young Stellar Objects (YSOs),
the compact H\,{\sc ii} region N\,81 and a similar object within N\,84, and
two red supergiants (RSGs). We use the spectra to constrain the presence and
temperature of cool dust and the excitation conditions within the neutral and
ionized gas, in the circumstellar environments and interfaces with the
surrounding interstellar medium. We compare these results with those obtained
in the LMC. The spectra of the sources in N\,81 (of which we also show the
ISO-LWS spectrum between 50--170 $\mu$m) and N\,84 both display strong
[O\,{\sc i}] $\lambda$63-$\mu$m and [O\,{\sc iii}] $\lambda88$-$\mu$m
fine-structure line emission. We attribute these lines to strong shocks and
photo-ionized gas, respectively, in a ``champagne flow'' scenario. The
nitrogen content of these two H\,{\sc ii} regions is very low, definitely
$N({\rm N})/N({\rm O})<0.04$ but possibly as low as $N({\rm N})/N({\rm
O})<0.01$. Overall, the oxygen lines and dust continuum are weaker in
star-forming objects in the SMC than in the LMC. We attribute this to the
lower metallicity of the SMC compared to that of the LMC. Whilst the dust mass
differs in proportion to metallicity, the oxygen mass differs less; both
observations can be reconciled with higher densities inside star-forming cloud
cores in the SMC than in the LMC. The dust in the YSOs in the SMC is warmer
(37--51 K) than in comparable objects in the LMC (32--44 K). We attribute this
to the reduced shielding and reduced cooling at the low metallicity of the
SMC. On the other hand, the {\it efficiency} of the photo-electric effect to
heat the gas is found to be indistinguishable to that measured in the same
manner in the LMC, $\approx0.1$--0.3\%. This may result from higher cloud-core
densities, or smaller grains, in the SMC. The dust associated with the two
RSGs in our SMC sample is cool, and we argue that it is swept-up interstellar
dust, or formed (or grew) within the bow-shock, rather than dust produced in
these metal-poor RSGs themselves. Strong emission from crystalline water ice
is detected in at least one YSO. The spectra constitute a valuable resource
for the planning and interpretation of observations with the {\it Herschel
Space Observatory} and the {\it Stratospheric Observatory For Infrared
Astronomy} (SOFIA).
\end{abstract}

%=========================================================================== 1
\section{Introduction}

About the cycle of gas and dust that drives galaxy evolution, much can be
learnt from the interfaces between the sources of feedback and the
interstellar medium (ISM), and between the ISM and the dense cores of
molecular clouds wherein new generations of stars may form. These regions are
characterized by the cooling ejecta from evolved stars and supernovae, and
clouds heated by the radiation and shocks from hot stars, in supernova
remnants and young stellar objects (YSOs) embedded in molecular clouds.

The interaction regions with the ISM lend themselves particularly well to
investigation in the infrared (IR) domain, notably in the 50--100 $\mu$m
region; cool dust ($\sim20$--100 K) shines brightly at these wavelengths, and
several strong atomic and ionic transitions of abundant elements (viz.\
[O\,{\sc i}] at $\lambda=63$ $\mu$m, [O\,{\sc iii}] at $\lambda=88$ $\mu$m,
and [N\,{\sc iii}] at $\lambda=57$ $\mu$m) provide both important diagnostics
of the excitation conditions and a mechanism for cooling. These diagnostic
signatures became widely accessible within the Milky Way, by virtue of the
{\it Kuiper Airborne Observatory} (KAO, see Erickson et al.\ 1984) and the
Long-Wavelength Spectrograph (LWS, Clegg et al.\ 1996) onboard the {\it
Infrared Space Observatory} (ISO, Kessler et al.\ 1996).

The gas-rich dwarf companions to the Milky Way, the Large and Small Magellanic
Clouds (LMC and SMC) offer a unique opportunity for a global assessment of the
feedback into the ISM and the conditions for star formation, something which
is much more challenging to obtain for the Milky Way due to our position
within it. The LMC and SMC are nearby ($d\sim50$ and 60 kpc, respectively:
Cioni et al.\ 2000; Keller \& Wood 2006) and already the scanning survey with
the {\it IR Astronomical Satellite} (IRAS) showed discrete sources of far-IR
emission in them (Schwering \& Israel 1989); it was used to describe the
diffuse cool dust and gas as well (Stanimirovi\'c et al.\ 2000). The stars,
star-forming regions, and ISM are also lower in metal content than similar
components of the Galactic Disc, $Z_{\rm LMC}\sim0.4$ Z$_\odot$ and $Z_{\rm
SMC}\sim0 1$--0.2 Z$_\odot$ (cf.\ discussion in Maeder, Grebel \& Mermilliod
1999). This offers the possibility to assess the effect metallicity has on the
dust content and on the heating and cooling processes, and to study these in
environments that are more similar to those prevailing in the early Universe
than the available Galactic examples (cf.\ Oliveira 2009).

The {\it Spitzer Space Telescope} marries superb sensitivity with exquisite
imaging quality, able to detect the far-IR emission from a significant
fraction of the total populations of YSOs, massive red supergiants (RSGs),
et cetera. The telescope also carried a facility, the MIPS-SED, to obtain
spectra at 50--100 $\mu$m, and we used this to target representative samples
of luminous 70-$\mu$m point sources in the LMC and SMC. The LMC spectra are
presented in paper I of this two-part series (van Loon et al.\ 2010); here we
present the results of the SMC observations and a comparison with the LMC
results.

%=========================================================================== 2
\section{Observations}

\begin{deluxetable*}{p{3mm}lllll}
\tabletypesize{\scriptsize}
\tablecaption{Description of compact sources in the SMC as targets for
MIPS-SED.}
\tablehead{
\colhead{\#}                                    &
\colhead{Principal name\ \tablenotemark{a}}     &
\colhead{Alternative name}                      &
\colhead{Object type\rlap{\ \tablenotemark{b}}} &
\colhead{RA and Dec (J2000)}                    &
\colhead{References}                            }
\startdata
%..............................................................................
1                        &
IRAS\,00429$-$7313       &                                               %jo2
                         &
YSO                      &
0 44 51.86 $-$72 57 34.2 &
36,42,48,55              \\
%..............................................................................
2                        &
IRAS\,00430$-$7326       &                                               %jo3
N\,10, LIN\,60           &
YSO                      &
0 44 56.30 $-$73 10 11.6 &
1,2,6,9,11,16,21,22,40,41,42,48,55,57 \\
%..............................................................................
3                        &
S3MC\,00464$-$7322\ \rlap{\tablenotemark{c}} &
IRAS\,00446$-$7339 ?     &                                               %jo6
YSO                      &
0 46 24.46 $-$73 22 07.3 &
4,16,27,30,42,46,47,48,55 \\
%..............................................................................
4                        &
GM\,103                  &
IRAS\,00486$-$7308, [GB98]\,S10 &
RSG                      &
0 50 30.62 $-$72 51 29.9 &
17,18,19,20,39,48,49,50,55 \\
%..............................................................................
5                        &
BMB-B\,75                &
                         &
RSG                      &
0 52 12.82 $-$73 08 52.8 &
3,7,16,50                \\
%..............................................................................
6                        &
S3MC\,00540$-$7321\ \rlap{\tablenotemark{d}} &
                         &                                               %jo17
YSO                      &
0 54 02.30 $-$73 21 18.7 &
42,55                    \\
%..............................................................................
7                        &
S3MC\,00541$-$7319\ \rlap{\tablenotemark{e}} &
                         &                                               %jo18
YSO                      &
0 54 03.36 $-$73 19 38.3 &
2,42,55                  \\
%..............................................................................
8                        &
S3MC\,01051$-$7159\ \rlap{\tablenotemark{f}} &
IRAS\,01035$-$7215 ?     &                                               %jo28
YSO                      &
1 05 07.25 $-$71 59 42.7 &
1,2,8,11,16,21,33,41,42,44,45,48,55 \\
%..............................................................................
9                        &
IRAS\,01039$-$7305       &
DEM\,S\,129, [MA93]\,1536 &
YSO                      &
1 05 30.22 $-$72 49 53.8 &
9,14,36,41,42,48,50,55,57,58 \\
%..............................................................................
10                       &
IRAS\,01042$-$7215       &
[GB98]\,S28              &
YSO                      &
1 05 49.30 $-$71 59 48.8 &
18,20,48,50,55           \\
%..............................................................................
11                       &
S3MC\,01070$-$7250\ \rlap{\tablenotemark{g}} &
IRAS\,01054$-$7307 ?     &                                               %jo30
YSO                      &
1 06 59.66 $-$72 50 43.1 &
2,9,41,42,55             \\
%..............................................................................
12                       &
N\,81                    &
IRAS\,01077$-$7327, DEM\,S\,138       &
H\,{\sc ii}              &
1 09 12.67 $-$73 11 38.4 &
1,2,6,9,10,11,12,13,15,16,21,22,23,24,\\
                         &
                         &
                         &
                         &
                         &
25,26,28,29,30,31,32,34,37,38,41,43,\\
                         &
                         &
                         &
                         &
                         &
48,51,52,53,54,55        \\
%..............................................................................
13                       &
S3MC\,01146$-$7318\ \rlap{\tablenotemark{h}} &
(in N\,84)               &                                               %jo31
YSO / H\,{\sc ii}        &
1 14 39.38 $-$73 18 29.3 &
2,5,16,35,41,42,55,56
%..............................................................................
\enddata
\tablenotetext{a}{Names of the type ``N\,[number]'' are
``LHA\,115-N\,[number]'' in full.}
\tablenotetext{b}{Used acronyms:
H\,{\sc ii} = region of ionized Hydrogen;
RSG         = Red Supergiant;
YSO         = Young Stellar Object.}
\tablenotetext{c}{Abbreviation of S3MC\,J004624.46$-$732207.30 (following the
IRAS convention).}
\tablenotetext{d}{Abbreviation of S3MC\,J005402.30$-$732118.70 (following the
IRAS convention).}
\tablenotetext{e}{Abbreviation of S3MC\,J005403.36$-$731938.30 (following the
IRAS convention).}
\tablenotetext{f}{Abbreviation of S3MC\,J010507.25$-$715942.70 (following the
IRAS convention).}
\tablenotetext{g}{Abbreviation of S3MC\,J010659.66$-$725043.10 (following the
IRAS convention).}
\tablenotetext{g}{Abbreviation of S3MC\,J011439.38$-$731829.26 (following the
IRAS convention).}
\tablerefs{
1.  Beasley et al.\ (1996);
2.  Bica \& Schmitt (1995);
3.  Blanco, McCarthy \& Blanco (1980);
4.  Bot et al.\ (2007);
5.  Bratsolis, Kontizas \& Bellas-Velidis (2004);
6.  Charmandaris, Heydari-Malayeri \& Chatzopoulos (2008);
7.  Cioni et al.\ (2003);
8.  Copetti (1990);
9.  Davies, Elliott \& Meaburn (1976);
10. Dennefeld \& Stasi\'nska (1983);
11. de Oliveira et al.\ 2000;
12. Dufour \& Harlow (1977);
13. Dufour, Shields \& Talbot (1982);
14. Evans et al.\ (2004);
15. Filipovi\'c et al.\ (1997);
16. Filipovi\'c et al.\ (2002);
17. Groenewegen (2004);
18. Groenewegen \& Blommaert (1998);
19. Groenewegen et al.\ (1995);
20. Groenewegen et al.\ (2000);
21. Henize (1956);
22. Henize \& Westerlund (1963);
23. Heydari-Malayeri, Le Bertre \& Magain (1988);
24. Heydari-Malayeri et al.\ (1999);
25. Heydari-Malayeri et al.\ (2002);
26. Heydari-Malayeri et al.\ (2003);
27. Hodge (1974);
28. Indebetouw, Johnson \& Conti (2004);
29. Israel \& Koornneef (1988);
30. Israel et al.\ (1993);
31. Kennicutt \& Hodge (1986);
32. Koornneef \& Israel (1985);
33. Kron (1956);
34. Krti\v{c}ka (2006);
35. Lindsay (1961);
36. Loup et al.\ (1997);
37. Mart\'{\i}n-Hern\'andez, Vermeij \& van der Hulst (2005);
38. Martins et al.\ (2004);
39. McSaveney et al.\ (2007);
40. Meynadier \& Heydari-Malayeri (2007);
41. Meyssonnier \& Azzopardi (1993);
42. Oliveira et al.\ (in prep.);
43. Pagel et al.\ (1978);
44. Pietrzy\'nski \& Udalski (1999);
45. Pietrzy\'nski et al.\ (1998);
46. Rubio, Lequeux \& Boulanger (1993);
47. Rubio et al.\ (1993b)
48. Schwering \& Israel (1989);
49. van Loon et al.\ (2001);
50. van Loon et al.\ (2008);
51. Vermeij \& van der Hulst (2002);
52. Vermeij et al.\ (2002);
53. Wilcots (1994a);
54. Wilcots (1994b);
55. Wilke et al.\ (2003);
56. Wisniewski \& Bjorkman (2006);
57. Wood et al.\ (1992);
58. Zijlstra et al.\ (1996).
}
\end{deluxetable*}

%------------------------------------------------------------------------- 2.1
\subsection{Data collection and processing}

\begin{figure*}
\epsscale{1.18}
\plotone{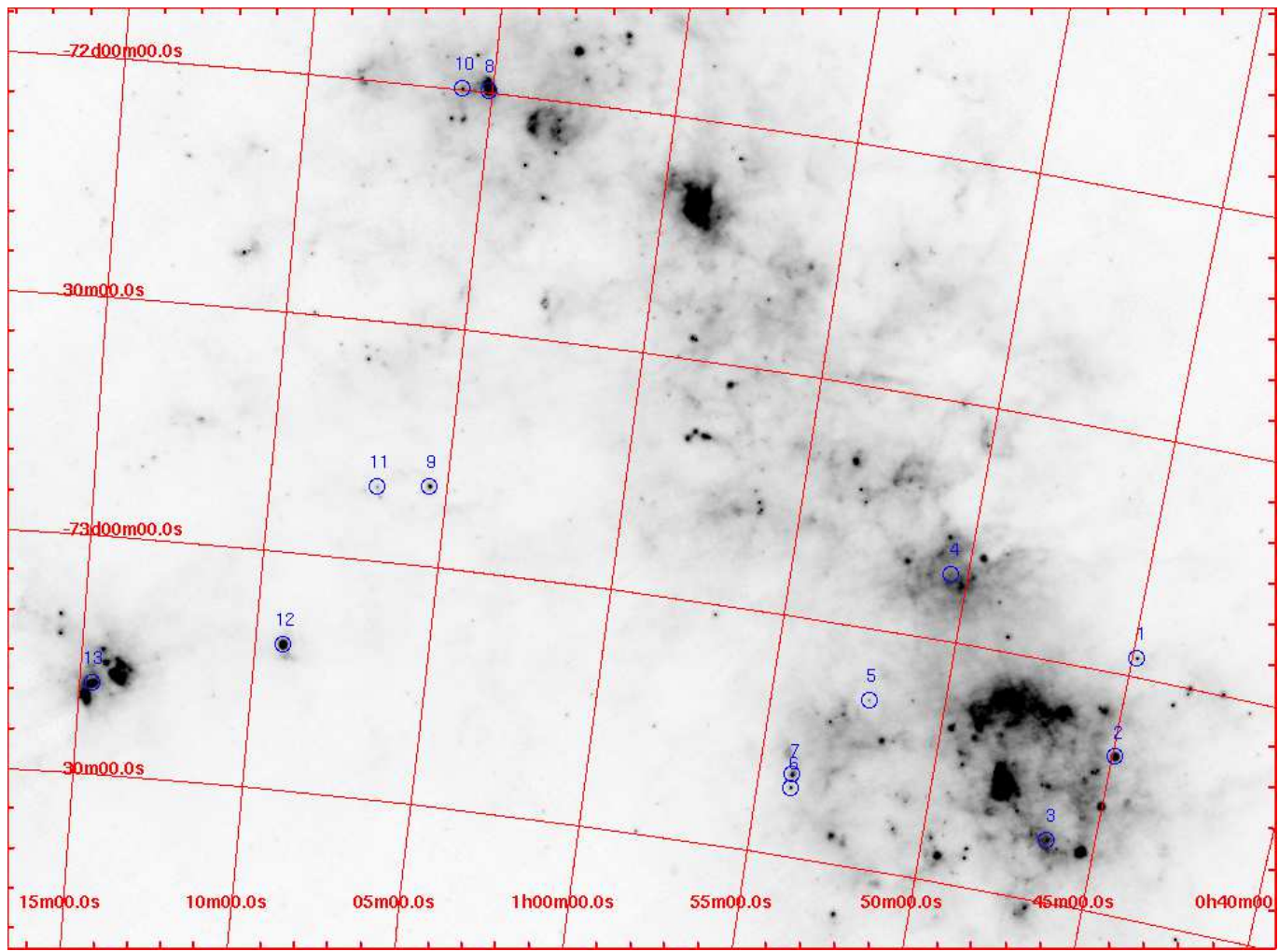}
\caption{All 13 MIPS-SED point sources plotted on top of the MIPS 70-$\mu$m
SAGE-SMC+S$^3$MC image. The main star-forming, dense-gas body of the Small
Magellanic Cloud stretches from North-East to South-West. The compact H\,{\sc
ii} region N\,81 (\#12) and H\,{\sc ii} regions N\,83+N\,84 (\#13) assume
rather isolated positions within the Shapley Wing (Shapley 1940), to the
East.}
\label{f1}
\end{figure*}

Our dataset comprises low-resolution spectra obtained using the Spectral
Energy Distribution (SED) mode of the {\it Multiband Imaging Photometer for
Spitzer} (MIPS; Rieke et al.\ 2004) onboard the {\it Spitzer Space Telescope}
(Werner et al.\ 2004), taken as part of the SMC-Spec program (PI: G.C.\
Sloan). The spectra cover $\lambda=52$--93 $\mu$m, at a spectral resolving
power $R\equiv\lambda/\Delta\lambda=15$--25 (two pixels) and a
cross-dispersion angular resolution of 13--24$^{\prime\prime}$ Full-Width at
Half-Maximum (sampled by $9.8^{\prime\prime}$ pixels). The slit is
$20^{\prime\prime}$ wide and $2.7^\prime$ long, but $0.7^\prime$ at one end of
the slit only covers $\lambda>65$ $\mu$m as a result of a dead readout. To
place the angular scales into perspective, $20^{\prime\prime}\equiv6$ pc at
the distance of the SMC. This is characteristic of a SNR, star cluster, or
molecular cloud core; it is smaller than a typical H\,{\sc ii} region, but
larger than a typical PN.

\begin{figure*}
\epsscale{1.1}
\plotone{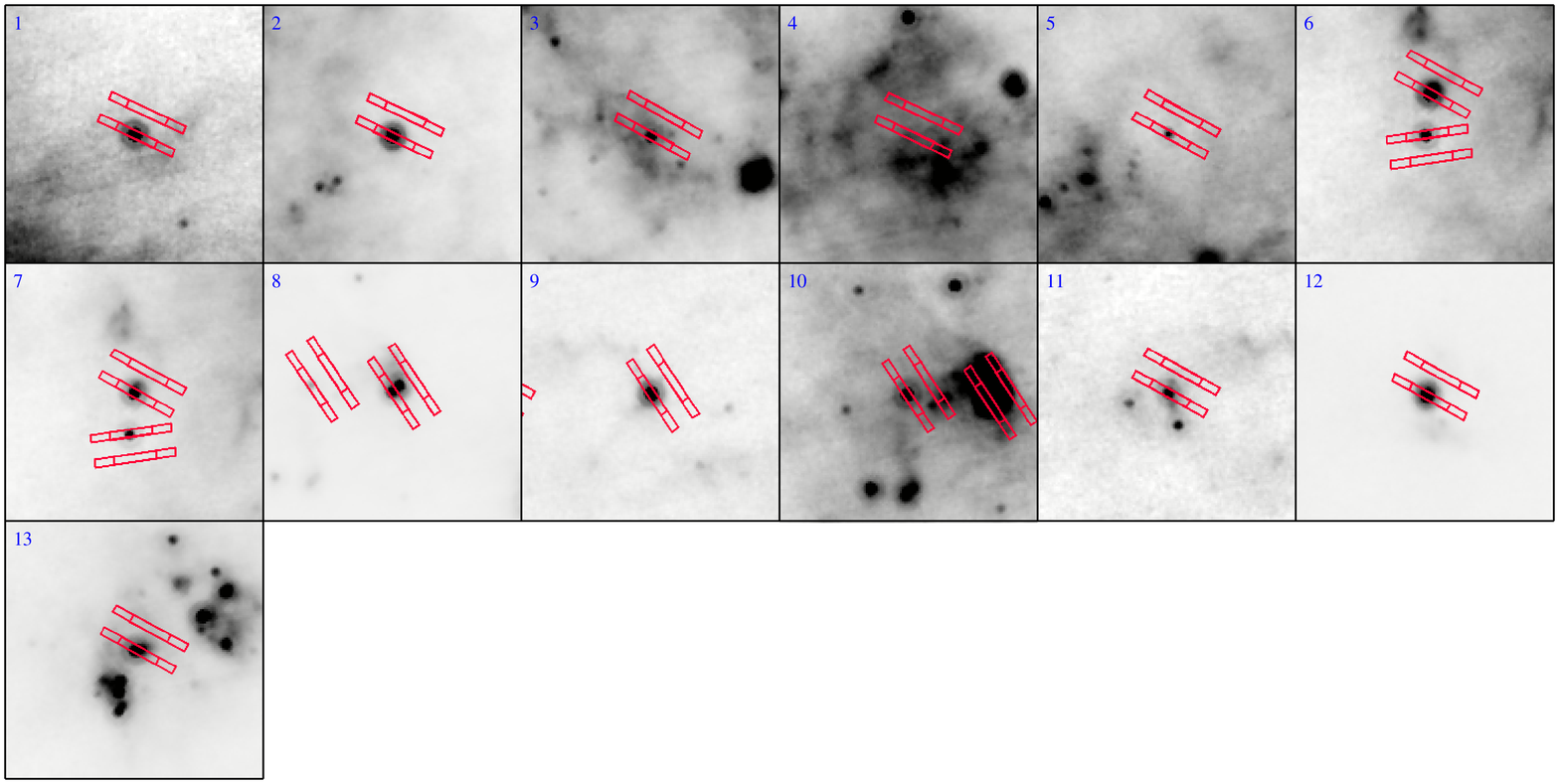}
\caption{Close-ups of the 70-$\mu$m emission centred on each of the 13
MIPS-SED targets, with overplotted the AOR footprints (on- and off-source slit
orientations). All images have North up and East to the left, and measure
$10^\prime$ on each side. The intensity scale is linear, but adjusted
individually such as to facilitate an assessment of the relative brightness of
the target compared to the background.}
\label{f2}
\end{figure*}

The target list, Table 1, is described in \S 2.2 and \S 3, and their
distribution on the sky is displayed in Fig.\ 1. The background spectrum was
measured at one of four possible chop positions, chosen to be free of other
discrete sources of 70-$\mu$m emission. This depends on the time of
observation. Fig.\ 2 shows 70-$\mu$m close-ups, extracted from the combined
SAGE-SMC {\it Spitzer} Legacy Program (Gordon et al.\ 2010, in preparation)
and S$^3$MC {\it Spitzer} survey (Bolatto et al.\ 2007), with the Astronomical
Observation Request (AOR) footprints overlain.

The raw data were processed with the standard pipeline version S16.1.1, and
the spectra were extracted and calibrated using the {\sc dat} software, v3.06
(Gordon et al.\ 2005). In some cases, small shifts in the centroids of
well-detected, unresolved spectral lines are noticeable; these always amount
to less than about a micrometer, i.e.\ within the spectral sampling, and no
attempt was made to correct for this (cf.\ figure 5 and its discussion in \S
5.1.2). Spectra were extracted from the on-off background-subtracted frame.
The extraction aperture was five pixels wide in the cross-dispersion
direction, and the (remaining) background level was determined in
a-few-pixel-wide apertures at either side of, and at some distance from, the
extraction aperture. The extracted spectrum was corrected to an infinite
aperture and converted to physical units, providing an absolute flux
calibrated spectrum (cf.\ Lu et al.\ 2008).

The uncertainties in the extracted spectrum stem mostly from inaccuracies in
the sky-subtraction. The statistical scatter was quantified, upon which the
errorbars (e.g., as plotted in Fig.\ 4) are based. However, larger deviations
may result from complex spatial structure of the sky emission, and this is not
possible to quantify. As this is important for an overall judgment of the
reliability of the spectrum, a ``quality'' flag was decided on the basis of a
subjective assessment of the reliability of the sky subtraction and spectrum
extraction. This is listed in Table 2, along with other descriptors of the
MIPS-SED data. Where possible in our analysis, we have accounted for measured
variations between adjacent spectral points in the computation of errorbars on
derived quantities, rather than to rely solely on the errorbars derived from
the statistical noise.

\begin{deluxetable}{p{3mm}lccc}
\tablecaption{Description of MIPS-SED data of compact sources in the SMC.}
\tablehead{
\colhead{\#}                             &
\colhead{AOR Key}                        &
\colhead{Integration\ \tablenotemark{a}} &
\colhead{Quality}                        &
\colhead{Extraction}                     }
\startdata
%..............................................................................
1                  &
27535360           &
\llap{6}\ $\times$\ \rlap{10} &
good               &
on--off            \\
%..............................................................................
2                  &
27535104           &
\llap{8}\ $\times$\ \rlap{3} &
good               &
on--off            \\
%..............................................................................
3                  &
27532288           &
\llap{7}\ $\times$\ \rlap{10} &
good               &
on--off            \\
%..............................................................................
4                  &
27536640           &
\llap{20}\ $\times$\ \rlap{10} &
ok                 &
on--off            \\
%..............................................................................
5                  &
27536384           &
\llap{20}\ $\times$\ \rlap{10} &
good               &
on--off            \\
%..............................................................................
6                  &
27532032           &
\llap{5}\ $\times$\ \rlap{10} &
good               &
on--off            \\
%..............................................................................
7                  &
27531776           &
\llap{3}\ $\times$\ \rlap{10} &
good               &
on--off            \\
%..............................................................................
8                  &
27531520           &
\llap{8}\ $\times$\ \rlap{3} &
good               &
on--off            \\
%..............................................................................
9                  &
27536128           &
\llap{3}\ $\times$\ \rlap{10} &
good               &
on--off            \\
%..............................................................................
10                 &
27535872           &
\llap{6}\ $\times$\ \rlap{10} &
good               &
on--off            \\
%..............................................................................
11                 &
27531264           &
\llap{20}\ $\times$\ \rlap{10} &
good               &
on--off            \\
%..............................................................................
12                 &
27535616           &
\llap{8}\ $\times$\ \rlap{3} &
good               &
on--off            \\
%..............................................................................
13                 &
27531008           &
\llap{3}\ $\times$\ \rlap{10} &
good               &
on--off
%..............................................................................
\enddata
\tablenotetext{a}{Total on-source integration time, $N_{\rm cycles}\times
t$(s).}
\end{deluxetable}

In our analysis of the MIPS-SED data we shall also make use of associated
photometry, from S$^3$MC at 24 and 70 $\mu$m with MIPS.

%------------------------------------------------------------------------- 2.2
\subsection{Target selection}

\begin{figure}
\epsscale{1.16}
\plotone{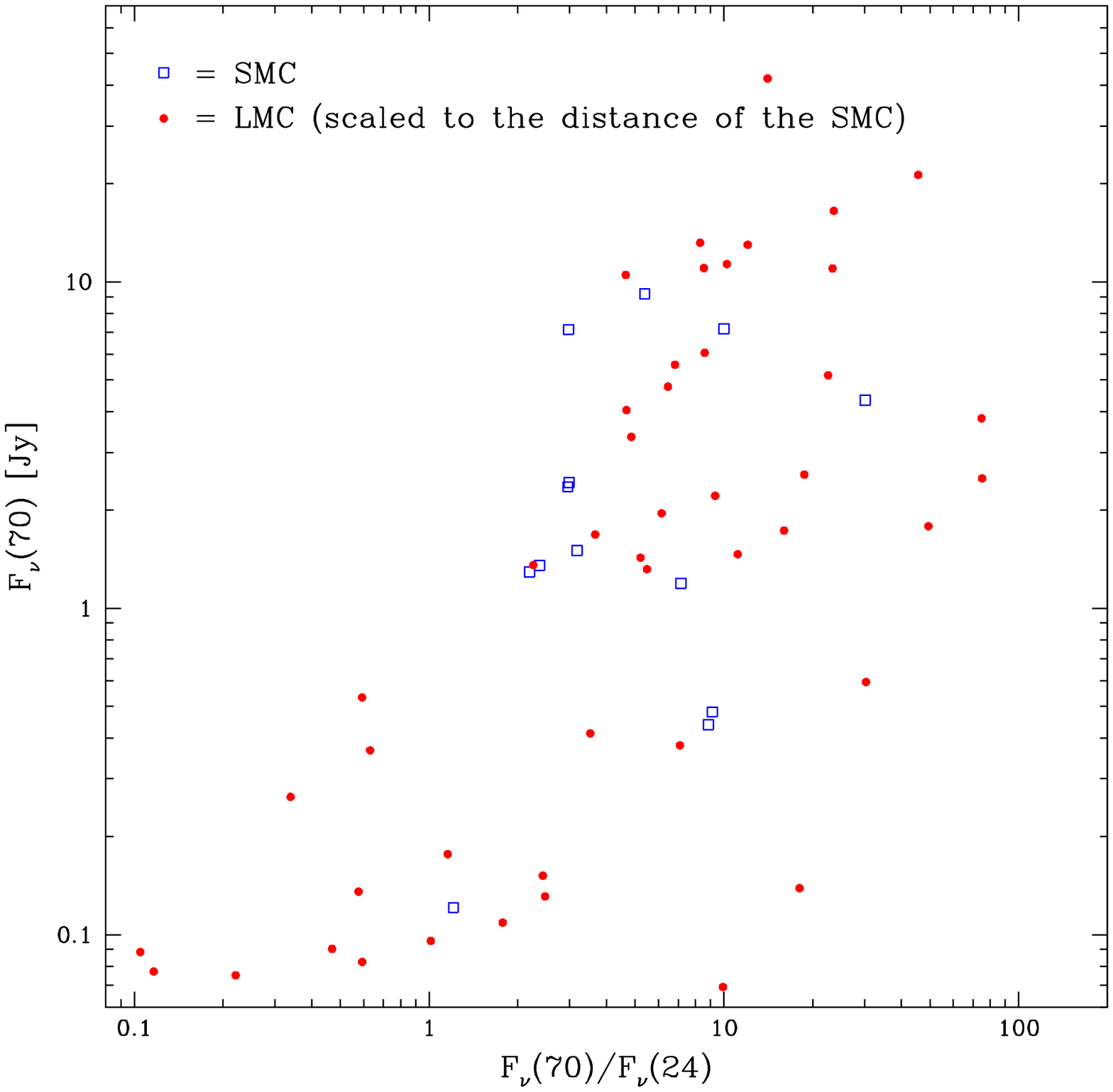}
\caption{$F_\nu(70)$ vs.\ $F_\nu(70)/F_\nu(24)$ diagram, with MIPS photometry
from the S$^3$MC catalog for the SMC-Spec MIPS-SED targets (squares). The
SAGE-Spec MIPS-SED targets in the LMC are also overplotted (dots).}
\label{f3}
\end{figure}

The targets were selected on the basis of the following criteria: (i) point
source appearance at 70 $\mu$m, and (ii) a minimum flux density at 70 $\mu$m
of $F_\nu(70)>0.3$ Jy. To reduce the large sample of potential targets to
within a reasonable time request, a further requirement was that at the time
of proposal submission there was a {\it Spitzer} IRS spectrum in the archive
or planned to be taken. Due to the limitations on observing time we could not
sample a wide range of object types and/or photometric properties (as we did
in the LMC sample), and we have deliberately restricted ourselves to generally
luminous YSOs and RSGs. They do, however, sample different regions within the
SMC (see Fig.\ 1). Table 3 summarizes the MIPS photometric properties of the
selected targets, and Fig.\ 3 shows them in the F$_\nu(70)$ vs.\
F$_\nu(70)$/F$_\nu(24)$ diagram in comparison to the LMC sample.

\begin{deluxetable}{p{3mm}lrrrr}
\tablecaption{MIPS photometric data of MIPS-SED targets in the SMC.}
\tablehead{
\colhead{\#}                       &
\colhead{AOR Target}               &
\colhead{$F_\nu(24)$}              &
\colhead{$\sigma(24)$}             &
\colhead{$F_\nu(70)$}              &
\colhead{$\sigma(70)$}             \\
                                   &
                                   &
[mJy]                              &
[mJy]                              &
[mJy]                              &
[mJy]                              }
\startdata
%..............................................................................
1                  &
IRAS\,00429$-$7313 &
588.27             &
0.20               &
1291               &
6                  \\
%..............................................................................
2                  &
IRAS\,00430$-$7326 &
717.69             &
0.54               &
7168               &
30                 \\
%..............................................................................
3                  &
S3MC\,00464$-$7322 &
166.67             &
0.10               &
1191               &
12                 \\
%..............................................................................
4                  &
GM\,103            &
100.00             &
0.07               &
121                &
8                  \\
%..............................................................................
5                  &
BMB-B\,75          &
52.55              &
0.06               &
481                &
4                  \\
%..............................................................................
6                  &
S3MC\,00540$-$7321 &
472.56             &
0.17               &
1502               &
6                  \\
%..............................................................................
7                  &
S3MC\,00541$-$7319 &
812.79             &
0.27               &
2423               &
11                 \\
%..............................................................................
8                  &
S3MC\,01051$-$7159 &
2402.03            &
0.55               &
7138               &
40                 \\
%..............................................................................
9                  &
IRAS\,01039$-$7305 &
796.12             &
0.24               &
2353               &
6                  \\
%..............................................................................
10                 &
IRAS\,01042$-$7215 &
569.61             &
0.14               &
1351               &
10                 \\
%..............................................................................
11                 &
S3MC\,01070$-$7250 &
49.71              &
0.07               &
440                &
2                  \\
%..............................................................................
12                 &
N\,81              &
1705.20            &
0.32               &
9190               &
41                 \\
%..............................................................................
13                 &
S3MC\,01146$-$7318 &
144.06             &
0.18               &
4340               &
45
%..............................................................................
\enddata
\end{deluxetable}

%=========================================================================== 3
\section{Comments on individual objects}

In the remainder of this paper, we shall refer to objects from the Henize
(1956) catalog as ``N\,[number]''; the full designation would be
``LHA\,115-N\,[number]''. Sources with only an S$^3$MC designation are
abbreviated following the IRAS convention (where the last digit of the RA part
derives from decimal minutes). Table 1 describes all MIPS-SED targets, with
literature references checked until Summer 2009.

Some targets have been studied before, and brief summaries of their nature are
given below. Essentially nothing was known previously about IRAS\,00429$-$7313
(\#1) and S3MC\,00540$-$7321 (\#6). All sources except BMB-B\,75 had already
been recognised as 170-$\mu$m point sources in Wilke et al.\ (2003). The
S$^3$MC sources were first identified as candidate YSOs by one of us (J.\
Oliveira); all YSOs in our sample have been confirmed to be YSOs with
ground-based IR observations and {\it Spitzer} IRS spectra (Oliveira et al.,
in prep.).

%------------------------------------------------------------------------- 3.1
\subsection{IRAS\,00430$-$7326 (\#2)}

The IR source is seen projected upon a nebulous cluster (Bica \& Schmitt 1995;
de Oliveira et al.\ 2000), and is embedded within the compact H\,{\sc ii}
region SMC-N\,10 (Henize 1956) = DEM\,S\,11 (Davies, Elliott \& Meaburn 1976);
Henize \& Westerlund (1963) estimated a nebular mass $\sim4$ M$_\odot$. It was
detected as a compact radio continuum source by Filipovi\'c et al.\ (2002).
Meynadier \& Heydari-Malayeri (2007) recognised it as a very-low excitation
``blob'', possibly powered by less massive stars; however, Charmandaris,
Heydari-Malayeri \& Chatzopoulos (2008) conclude that the mid-IR properties
resemble simply a scaled-down version from more prominent H\,{\sc ii} regions.

%------------------------------------------------------------------------- 3.2
\subsection{S3MC\,00464$-$7322 (\#3)}

This IR object is possibly associated with the dark cloud \#7 in Hodge (1974),
which was also the strongest $^{12}$CO(1--0) and $^{13}$CO(1--0) detection in
the ESO/SEST key programme (Israel et al.\ 1993), SMC-B1, which is split up
into SMC-B1\,2 and 3 each containing an estimated few $10^4$ M$_\odot$ (Rubio,
Lequeux \& Boulanger 1993; Rubio et al.\ 1993b). The mm-continuum emission
from B1\,2 was much brighter and more extended than that from B1\,3 (Bot et
al.\ 2007).

%------------------------------------------------------------------------- 3.3
\subsection{GM\,103 (\#4)}

GM\,103 is identified with IRAS\,00486$-$7308, near N\,36 (Henize 1956). A
ground-based N-band spectrum showed the 10-$\mu$m silicate feature in emission
(Groenewegen et al.\ 1995); modelling of the SED yielded a mass-loss rate of
$\dot{M}\sim10^{-5}$ M$_\odot$ yr$^{-1}$ and a luminosity near the AGB maximum
associated with the Chandrasekhar limit for the core mass. Groenewegen \&
Blommaert (1998) presented an optical spectrum, of late-M spectral type;
apparently cooler than the M\,4 type which Groenewegen has assumed in other
works. McSaveney et al.\ (2007) presented an analysis of the pulsation
properties. They confirmed the long pulsation period of $P=1070$ d, at a large
near-IR amplitude of $\Delta K=1.3$ mag, and a luminosity near the AGB
maximum. From pulsation modelling they derived a (current) mass of $M_{\rm
puls}\approx 6$ M$_\odot$. Van Loon et al.\ (2008) presented a 3--4 $\mu$m
spectrum, resembling that of a cool luminous star.

%------------------------------------------------------------------------- 3.4
\subsection{BMB-B\,75 (\#5)}

An M\,6-type star (Blanco, McCarthy \& Blanco 1980), it has a luminosity near
the AGB maximum, and shows Mira-type variability with a very long period of
$P=1453$ d (Cioni et al.\ 2003). A 3--4 $\mu$m spectrum was presented in van
Loon et al.\ (2008); it resembles that of a cool giant star. There is a
compact radio continuum source within a few arcsec, which was only detected at
2.37 GHz (not at 1.42, 4.80 or 8.64 GHz), at a level of 8 mJy (Filipovi\'c et
al.\ 2002).

%------------------------------------------------------------------------- 3.5
\subsection{S3MC\,00541$-$7319 (\#7)}

This source is seen projected against the star cluster [H86]\,144 (Bica \&
Schmitt 1995).

%------------------------------------------------------------------------- 3.6
\subsection{S3MC\,01051$-$7159 (\#8)}

This IR source sits in the middle of the cluster OGLE-CL\,SMC\,147
(Pietrzy\'nski et al.\ 1998), which has an estimated age of $t\approx
13^{+12}_{-7}$ Myr (Pietrzy\'nski \& Udalski 1999). There are several compact
H\,{\sc ii} regions nearby, the closest of which is \#1520 in Meyssonnier \&
Azzopardi (1993), at $6^{\prime\prime}$. These sources form part of the much
larger, loose stellar association NGC\,395 (Kron 1956; Bica \& Schmitt 1995)
associated with N\,78 (Henize 1956) = DEM\,S\,126 (Davies et al.\ 1976), which
contains a further two compact H\,{\sc ii} regions (Indebetouw, Johnson \&
Conti 2004) and is an extended source of radio continuum emission (Filipovi\'c
et al.\ (2002). Curiously, at $7^{\prime\prime}$, opposite to [MA93]\,1520, is
the background galaxy 2MFGC\,779 (Mitronova et al.\ 2004).

%------------------------------------------------------------------------- 3.7
\subsection{IRAS\,01039$-$7305 (\#9)}

This red IR object is a compact source of H$\alpha$ emission, \#1536 in
Meyssonnier \& Azzopardi (1993), identified with DEM\,S\,129 (Davies et al.\
1976), and probably an early-B type stellar object (\#2027 in Evans et al.\
2004). It was recognized as a YSO for its Br$\alpha$ and Pf$\gamma$ emission
lines on top of a red continuum in the 3--4 $\mu$m spectrum (van Loon et al.\
2008).

%------------------------------------------------------------------------- 3.8
\subsection{IRAS\,01042$-$7215 (\#10)}

This very red IR source (Groenewegen \& Blommaert 1998) has long been
considered as a candidate dust-enshrouded AGB star or RSG. Groenewegen et al.\
(2000) modelled the SED assuming a spectral type of M\,8, but the obtained
luminosity is rather low ($10^3$ L$_\odot$). This raises doubts about it being
a cool, evolved star. Indeed, it was recognized as a YSO by  van Loon et al.\
(2008), on the basis of water ice absorption and Br$\alpha$ and Pf$\gamma$
emission lines on a very red continuum in the 3--4 $\mu$m spectrum.

%------------------------------------------------------------------------- 3.9
\subsection{S3MC\,01070$-$7250 (\#11)}

This source can be identified with a faint, compact H$\alpha$ emission-line
object, \#1607 in Meyssonnier \& Azzopardi (1993), which is situated in a
stellar association within the H\,{\sc ii} region DEM\,S\,133 (Davies et al.\
1976; Bica \& Schmitt 1995).

%------------------------------------------------------------------------ 3.10
\subsection{N\,81 (\#12)}

The IR source IRAS\,01077$-$7327 is associated with the rather isolated, very
bright compact H\,{\sc ii} region N\,81 (Henize 1956) = DEM\,S\,138 (Davies et
al.\ 1976). A comprehensive {\it HST} optical imaging and
spectroscopic-imaging study was performed by Heydari-Malayeri et al.\ (1999,
2002), who identified several O-type stars within a few arcseconds in the core
of the H\,{\sc ii} region. These stars are young, possible pre-main sequence,
inferred from their low luminosity and weak stellar winds. Nonetheless, they
induce a ``Champagne flow'' (Heydari-Malayeri et al.\ 1999), with a
shock-compressed front in one direction (West), and an ionized tail in the
opposite direction (East). This picture was confirmed with high-resolution
radio continuum images, by Indebetouw, Johnson \& Conti (2004) and
Mart\'{\i}n-Hern\'andez, Vermeij \& van der Hulst (2005). The latter estimated
an ionized mass of $M_{\rm ion}\approx 56$ M$_\odot$, rather less than Henize
\& Westerlund (1963), who had estimated $M_{\rm ion}\approx 660$ M$_\odot$
from optical spectroscopy. The detection of H$_2$ in its excited S1(1--0)
transition at $\lambda=2.121$ $\mu$m (Koornneef \& Israel 1985) confirmed the
presence of a mild shock. CO was also detected (Israel et al.\ 1993), though
weakly. Vermeij et al.\ (2002) presented {\it ISO} SWS and LWS spectra, and we
compare with their measurements (see \S 5).

%------------------------------------------------------------------------ 3.11
\subsection{S3MC\,01146$-$7318 (\#13)}

The nature of this source is unclear. It is seen projected against the rich
star cluster NGC\,460 (Kron 1956) = IRAS\,01133$-$7333 (Loup et al.\ 1997),
with a rather uncertain age estimated to be $t\sim20(\pm14)\times10^6$ yr
(Hodge 1983). The cluster has plenty of (candidate) Be stars (Wisniewski \&
Bjorkman 2006), yet the nearest recorded objects are $6^{\prime\prime}$ away.
One of these is an emission-line star, \#507 in Lindsay (1961) = \#1792 in
Meyssonnier \& Azzopardi (1993), which is a candidate Be star (Wisniewski \&
Bjorkman 2006). Unrelated to that source but not much further away
(8--$14^{\prime\prime}$) is an extended radio continuum source (Filipovi\'c et
al.\ 2002). The cluster is embedded in the bright H\,{\sc ii} region N\,84A
(Henize 1956) = DEM\,S\,151 (Davies et al.\ 1976), for which Copetti (1990)
estimated an ionized mass of $M_{\rm ion}\approx1200$ M$_\odot$. Testor \&
Lortet (1987), pioneering CCD imaging, argued for an extended period of
sequential star formation in this part of the SMC, and the presence of
unevolved O stars in the N\,84 region was taken as a sign of continued star
formation. Weak $^{12}$CO(1--0) emission appears to be associated with it,
though the bulk of emission arises from further North-West (Mizuno et al.\
2001). Hodge (1974) identified a dark cloud in this region. More detailed
investigations of the molecular and dust distribution in the N\,83/N\,84
region were conducted by Bolatto et al.\ (2003) and Lee et al.\ (2009), but
S3MC\,01146$-$7318 does not appear to be associated with any conspicuous
feature in their maps.

%=========================================================================== 4
\section{Results}

\begin{figure*}
\epsscale{1.17}
\plotone{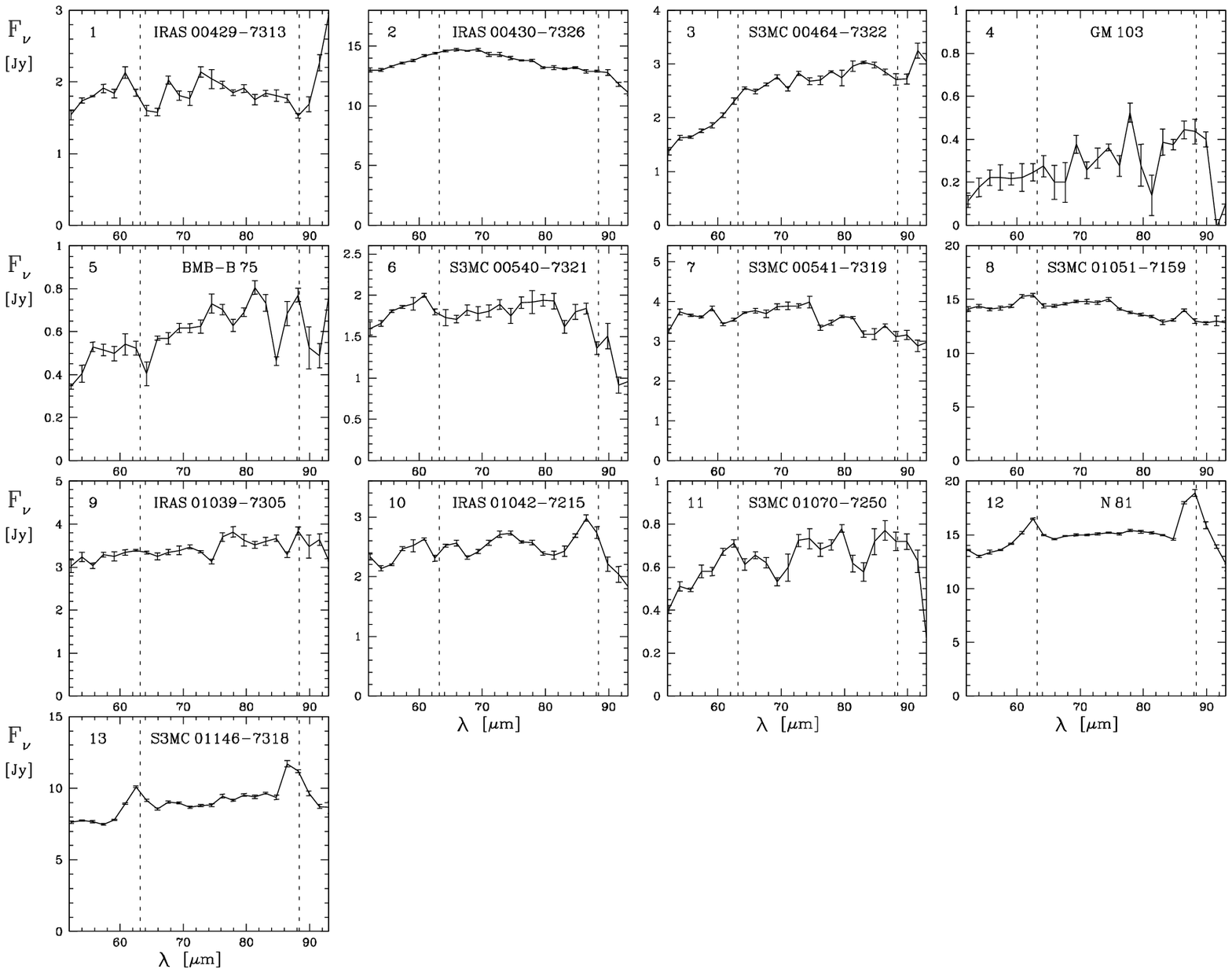}
\caption{MIPS-SED spectra of all 13 targets in the SMC. Vertical dashed lines
indicate the positions of the [O\,{\sc i}] and [O\,{\sc iii}] fine-structure
emission lines at $\lambda=63$ and 88 $\mu$m, respectively.}
\label{f4}
\end{figure*}

The MIPS-SED spectra of all 13 targets are presented in Fig.\ 4. All targets
were considerably brighter than the surrounding sky emission, except GM\,103
(of which nevertheless a reliable spectrum could be extracted). Table 4
summarizes properties derived from the {\it Spitzer} data.

One or two fine-structure emission lines may be seen, [O\,{\sc i}] $^3$P(1--2)
and [O\,{\sc iii}] $^3$P(0--1), at $\lambda=63.2$ and 88.4 $\mu$m,
respectively. These are discussed in \S 5.1. There is no convincing detection
of the [N\,{\sc iii}] $^2$P(1/2--3/2) transition at $\lambda=57.3$ $\mu$m, and
we discuss the implication in \S 5.2. There is evidence for additional
discrete features in the spectra of some objects, but their identification is
uncertain. They are discussed in \S 5.4. The slope of the continuum is an
indication of the dust temperature and is discussed in \S 5.3.

\begin{deluxetable*}{p{3mm}llclccrl}
\tablecaption{Quantities derived from the MIPS-SED spectra of compact sources
in the SMC.}
\tablehead{
\colhead{\#}                                     &
\colhead{AOR Target}                             &
\colhead{Type\ \tablenotemark{a}}                &
\colhead{$\alpha$\ \tablenotemark{b}}            &
\colhead{$T_{\rm dust}$\ \tablenotemark{c}}      &
\colhead{$L({\rm [O\,I]})$\ \tablenotemark{d}}   &
\colhead{$L({\rm [O\,III]})$\ \tablenotemark{d}} &
\colhead{$L({\rm FIR})$\ \tablenotemark{e}}      &
\colhead{Notes}                                  \\
                                                 &
                                                 &
                                                 &
                                                 &
[K]                                              &
\multicolumn{3}{c}{------------------------\ \ \ [L$_\odot$]\ \ \ ------------------------} &
                                            }
\startdata
%..............................................................................
1                  &
IRAS\,00429$-$7313 &
F0                 &
\llap{$-$}$0.007\pm0.021$   &
$48_{-1}^{+1}$     &
$69\pm23$          &
$<1$\rlap{5}       &
$1.814_{+0.04}^{-0.038}\times10^4$ &
                   \\
%..............................................................................
2                  &
IRAS\,00430$-$7326 &
F0                 &
\llap{$-$}$0.022\pm0.010$ &
$48_{-1}^{+1}$     &
$67\pm31$          &
$80\pm25$          &
$1.323_{+0.031}^{-0.028}\times10^5$ &
f                  \\
%..............................................................................
3                  &
S3MC\,00464$-$7322 &
C1                 &
$0.673\pm0.020$    &
$37_{-1}^{+1}$     &
$55\pm17$          &
$<1$\rlap{7}       &
$3.06_{+0.18}^{-0.16}\times10^4$ &
f?                 \\
%..............................................................................
4                  &
GM\,103            &
C0                 &
$0.78\pm0.06$      &
$35_{-1}^{+1}$     &
$<1$\rlap{9}       &
\llap{5}$4\pm7$    &
$4.44_{+0.31}^{-0.27}\times10^3$ &
                   \\
%..............................................................................
5                  &
BMB-B\,75          &
C0                 &
$0.32\pm0.11$      &
$42_{-2}^{+2}$     &
$<1$\rlap{4}       &
$41\pm10$          &
$6.2_{+0.5}^{-0.4}\times10^3$ &
                   \\
%..............................................................................
6                  &
S3MC\,00540$-$7321 &
F0                 &
\llap{$-$}$0.016\pm0.036$ &
$48_{-1}^{+1}$     &
$<2$\rlap{3}       &
$47\pm18$          &
$1.770_{+0.04}^{-0.037}\times10^4$ &
f?                 \\
%..............................................................................
7                  &
S3MC\,00541$-$7319 &
W0b                &
\llap{$-$}$0.147\pm0.017$ &
$51_{-1}^{+1}$     &
$<1$\rlap{8}       &
$43\pm22$          &
$3.34_{+0.06}^{-0.05}\times10^4$ &
                   \\
%..............................................................................
8                  &
S3MC\,01051$-$7159 &
W2b                &
\llap{$-$}$0.080\pm0.018$ &
$50_{-1}^{+1}$     &
\llap{2}$87\pm56$  &
$51\pm45$          &
$1.341_{+0.026}^{-0.023}\times10^5$ &
                   \\
%..............................................................................
9                  &
IRAS\,01039$-$7305 &
C0b                &
$0.121\pm0.029$    &
$45_{-1}^{+1}$     &
$47\pm28$          &
$<2$\rlap{5}       &
$3.54_{+0.11}^{-0.10}\times10^4$ &
                   \\
%..............................................................................
10                 &
IRAS\,01042$-$7215 &
F3b                &
$0.21\pm0.05$      &
$44_{-1}^{+1}$     &
$<2$\rlap{3}       &
$63\pm17$          &
$2.66_{+0.09}^{-0.08}\times10^4$ &
                   \\
%..............................................................................
11                 &
S3MC\,01070$-$7250 &
C1                 &
$0.32\pm0.07$      &
$42_{-1}^{+1}$     &
\llap{2}$2\pm7$    &
\llap{1}$4\pm9$    &
$6.84_{+0.27}^{-0.24}\times10^3$ &
                   \\
%..............................................................................
12                 &
N\,81              &
F2                 &
$0.21\pm0.05$      &
$44_{-1}^{+1}$     &
\llap{5}$18\pm27$  &
\llap{7}$48\pm38$  &
$1.56_{+0.05}^{-0.05}\times10^5$ &
                   \\
%..............................................................................
13                 &
S3MC\,01146$-$7318 &
C2b                &
$0.35\pm0.05$      &
$41_{-1}^{+1}$     &
\llap{5}$39\pm22$  &
\llap{4}$05\pm28$  &
$1.038_{+0.04}^{-0.039}\times10^5$ &

%..............................................................................
\enddata
\tablenotetext{a}{The spectral classification scheme is described in \S 4.2.}
\tablenotetext{b}{The spectral slope is defined as: $\alpha \equiv 2.44\
\frac{F_\nu(85)-F_\nu(55)}{F_\nu(85)+F_\nu(55)}$.}
\tablenotetext{c}{The dust temperature was estimated from the spectral slope
as described in \S 5.3.1.}
\tablenotetext{d}{Upper limits to the line luminosities correspond to
1-$\sigma$ values.}
\tablenotetext{e}{The far-IR luminosity was estimated based on the dust
temperature as described in Paper I, assuming a distance of 60 kpc.}
\tablenotetext{f}{Possible broad crystalline water ice band between
$\lambda=60$--70 $\mu$m.}
\end{deluxetable*}

%------------------------------------------------------------------------- 4.1
\subsection{Clarification of the nature of S3MC\,01146$-$7318}

The MIPS-SED spectrum of this object (Fig.\ 4, \#13) looks very much like that
of N\,81 (Fig.\ 4, \#12), so it is likely an (ultra)compact H\,{\sc ii} region
too. Although N\,84 has been suggested to be a somewhat evolved molecular
cloud complex, it is not unprecedented to encounter YSO-like objects near more
mature massive stars, e.g., the proplyd $\sigma$\,Ori-IRS1 (van Loon \&
Oliveira 2003). The identification of this source in N\,84 suggests that star
formation is still on-going there.

%------------------------------------------------------------------------- 4.2
\subsection{Classification of the MIPS-SED spectra}

A simple classification scheme based on the spectral appearance, ``The Keele
System'', was first introduced in Paper I. The primary spectral type is
defined as follows:
\begin{itemize}
\item[$\bullet$]{An upper-case letter denotes the continuum slope, for a
spectrum expressed in F$_\nu$ as a function of $\lambda$: C = rising (e.g.,
relatively cold dust); F = flat (this includes spectra with a peak mid-way the
MIPS-SED range); W = declining (e.g., relatively warm dust);}
\item[$\bullet$]{Following the upper-case letter, a number denotes the
presence of the [O\,{\sc i}] and [O\,{\sc iii}] lines: 0 = no oxygen lines are
present; 1 = the [O\,{\sc i}] line is present, but the [O\,{\sc iii}] line is
not; 2 = both [O\,{\sc i}] and [O\,{\sc iii}] lines are present; 3 = the
[O\,{\sc iii}] line is present, but the [O\,{\sc i}] line is not.}
\end{itemize}
A secondary classification is based on additional features: a lower-case
letter ``b'' may follow the primary type in the presence of a bump in the
$\lambda\sim70$--80 $\mu$m region.

We have classified all spectra (Table 4), erring on the side of caution with
respect to the detection of spectral lines. So, for instance, an object with
spectral type C0 may still display weak oxygen lines in a higher-quality
spectrum.

%=========================================================================== 5
\section{Discussion}

We first discuss the oxygen fine-structure emission lines (\S 5.1), followed
by the nitrogen line (\S 5.2), and then the dust continuum (\S 5.3) and
discrete features possibly due to ice, molecules, or minerals (\S 5.4). At the
end (\S 5.5), we summarize the population MIPS-SED characteristics and compare
with those in the LMC presented in Paper~I.

%------------------------------------------------------------------------- 5.1
\subsection{Oxygen}

The diagnostic value of the oxygen lines was described in detail in Paper I
(cf.\ Tielens \& Hollenbach 1985). We recall that the [O\,{\sc i}] line at
$\lambda=63$ $\mu$m is an important cooling line in relatively dense and
neutral or weakly-ionized gas, and it is enhanced if slow shocks are present.
The [O\,{\sc iii}] line at $\lambda=88$ $\mu$m is a measure of the electron
density in ionized gas such as that which occupies H\,{\sc ii} regions.

The line luminosities in the SMC sources can be found in Table 4, for an
assumed distance of 60 kpc. They were computed in identical fashion to those
of the LMC sources in Paper I, by summing the three spectral points centered
on the line (data registered at 60.81, 62.52 and 64.23 $\mu$m for the [O\,{\sc
i}] line, and at 86.46, 8817 and 89.88 $\mu$m for the [O\,{\sc iii}] line)
after subtracting a continuum obtained by linear interpolation between the
spectral points immediately adjacent to the integration interval. The error
was computed by adding in quadrature the errors in the three spectral points
that were summed, and three times the error in the mean of the two continuum
anchors (to account for the error in the continuum estimate at each of the
three spectral points).

%....................................................................... 5.1.1
\subsubsection{Oxygen in star-forming regions and YSOs}

The oxygen lines in the objects associated with star formation are weak. In
fact, on closer inspection, the only really convincing detections are in
S3MC\,01051$-$7159, N\,81, and S3MC\,01146$-$7318 (Fig.\ 4, \#8, \#12 and
\#13), where both the [O\,{\sc i}] and [O\,{\sc iii}] lines are clearly
discernible. In the LMC sources associated with star formation, [O\,{\sc i}]
was more often detected; this is partly because the detection threshold (in
L$_\odot$) was about half that in the SMC observations, but most LMC sources
were above the SMC threshold too.

%....................................................................... 5.1.2
\subsubsection{Shocks and ionized gas in the compact H\,{\sc ii} regions N\,81
and S3MC\,01146$-$7318 in N\,84}

In the case of N\,81, the MIPS-SED aperture comfortably includes the entire
radio source associated with this H\,{\sc ii} region. Previous observations
made by Vermeij et al.\ (2002) using the ISO-LWS employed a much larger
aperture, with a circular $80^{\prime\prime}$ diameter; there is not much more
emission than captured in the MIPS-SED slit and thus the spectra are
comparable. Vermeij et al.\ measured $F({\rm [O\,I]})=37\times10^{-16}$ W
m$^{-2}$ and $F({\rm [O\,III]})=74\times10^{-16}$ W m$^{-2}$, or 414 and 829
L$_\odot$, respectively, which is not radically different from the line fluxes
we determined (Table 4).

\begin{figure}
\epsscale{1.17}
\plotone{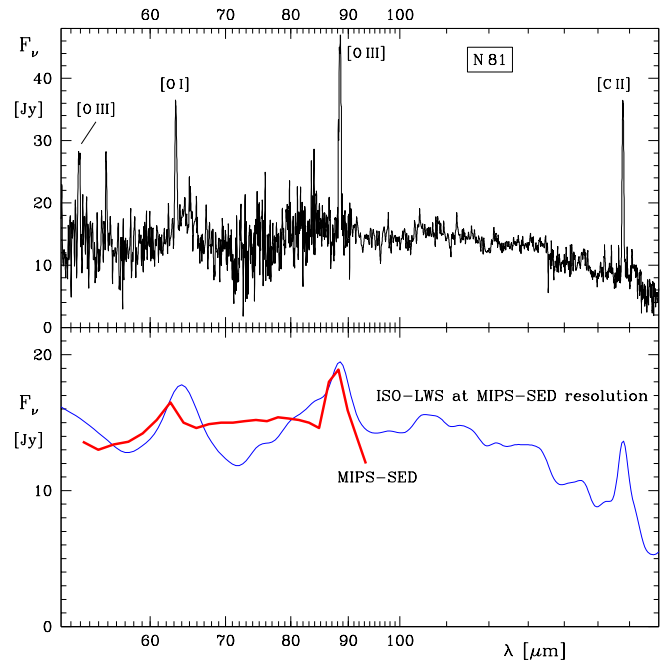}
\caption{{\it Top}: ISO-LWS spectrum of N\,81. {\it Bottom}: Comparison
between the MIPS-SED spectrum and the ISO-LWS spectrum after convolving the
latter with the MIPS-SED instrumental profile.}
\label{f5}
\end{figure}

Vermeij et al.\ did not publish the ISO-LWS spectrum, so we extracted the
pipeline-processed spectrum from the ISO Archive. It is composed of individual
segments, with large jumps in flux level from one segment to the other. We
used the overlap regions to calibrate them relative to one another, and
applied a global correction factor to bring the flux level in line with that
in our MIPS-SED spectrum. The result is shown in Fig.\ 5, both at the ISO-LWS
spectral resolution and convolved with a Gaussian of width 3.5 $\mu$m
(MIPS-SED spectral resolution). The continuum is more reliably measured in the
MIPS-SED spectrum; the oxygen lines are better resolved in the ISO-LWS
spectrum. The [O\,{\sc iii}] $\lambda$88-$\mu$m line was recorded in two
overlapping LWS segments but at rather different intensities; nevertheless the
convolved line profile matches that in the MIPS-SED spectrum quite well. The
[O\,{\sc i}] $\lambda$63-$\mu$m line is more discrepant, but this is due to a
bump in the ISO-LWS spectrum longward of the oxygen line --- it could be due
to water ice, but the corresponding ISO-LWS segment drops suspiciously at that
side, and it was not recorded in the MIPS-SED spectrum.

The ISO-LWS spectrum of N\,81 (Fig.\ 5) also includes the [O\,{\sc iii}]
$\lambda$52-$\mu$m line, and the important [C\,{\sc ii}] $\lambda$158-$\mu$m
cooling line. The latter is probably sampled from a larger region than that
sampled by the MIPS-SED; the [O\,{\sc i}] $\lambda$63-$\mu$m line is expected
to dominate the cooling in the denser inner part of the cloud (see \S 5.5.2).

The strength of the oxygen lines of such different ionization stage indicates
the coincidence of both ionized gas and shocks. This is consistent with the
idea of the ``champagne flow'' (note that in this scenario there is a distinct
spatial segregation between the shocked gas and the photo-ionized gas, but
this is not resolved in our observations). The detection of the H$_2$ S1(1--0)
line at 2.121 $\mu$m in N\,81 (Koornneef \& Israel 1985) is also indicative of
a mild shock.

The source S3MC\,01146$-$7318, in N\,84, is virtually identical to N\,81, only
a little bit fainter. We thus conjecture that it is of a similar nature, i.e.\
a compact H\,{\sc ii} region with a champagne flow structure. Stars must
recently have formed here, and are now blowing the cloud apart through their
mechanical and radiative feedback.

%....................................................................... 5.1.3
\subsubsection{Oxygen in evolved objects}

No oxygen line emission is seen in either GM\,103 or BMB-B\,75 --- the
apparently significant detection of [O\,{\sc iii}] is an anomaly caused by the
poor signal near the edge of the MIPS-SED range (these are two of the faintest
targets). This is not very surprising as RSGs lack a potent excitation
mechanism for these lines, and even in the LMC the [O\,{\sc i}] line was
tentatively detected only in the most luminous RSG, WOH\,G064 (Paper I).

%------------------------------------------------------------------------- 5.2
\subsection{Nitrogen}

Evidence has been presented of a low nitrogen content in relation to the
abundance of other common metals such as oxygen, if the overall metallicity is
low (Rubin et al.\ 1988; Simpson et al.\ 1995; Roelfsema et al.\ 1998; Rudolph
et al.\ 2006). This hinges on evidence in the Magellanic Clouds, which is
rather limited. Any additional or refined measurements in the Magellanic
Clouds are most welcome to solidify --- or weaken --- the observed trend.

In Paper I we used the best MIPS-SED spectra of LMC compact H\,{\sc ii}
regions to estimate that the nitrogen-to-oxygen ratio, $N({\rm N})/N({\rm
O})_{\rm LMC}\la0.1$ and possibly as low as $\la0.03$, which is much below the
0.1--0.4 typical in the Galactic Disc. Like in the LMC sample, no nitrogen
line is detected in the SMC sample. Can we still place interesting upper
limits? This depends on the strength of the reference oxygen line, preferably
the [O\,{\sc iii}] line at 88 $\mu$m as the ionization potential of the lower
level of this transition is similar to that of the [N\,{\sc iii}] line at 57
$\mu$m. The [O\,{\sc iii}] is not phenomenally bright in any SMC source and we
are anticipating $N({\rm N})/N({\rm O})_{\rm SMC}<0.05$. But the spectra of
N\,81 and S3MC\,01146$-$7318 are very good, displaying strong [O\,{\sc iii}]
lines.

Assuming a detection level of the [N\,{\sc iii}] line at 57 $\mu$m similar to
the 1-$\sigma$ noise level near the [O\,{\sc i}] line at 63 $\mu$m, we would
get [N\,{\sc iii}]/[O\,{\sc iii}]$\la0.04$ in N\,81 and $\la0.05$ in
S3MC\,01146$-$7318. To place this improvement in context, previous data on
N\,81 obtained with ISO-LWS by Vermeij et al.\ (2002) yielded a limit of
$F({\rm [N\,III]})<38\times10^{-16}$ W m$^{-2}$, or 426 L$_\odot$, and thus
(see above) [N\,{\sc iii}]/[O\,{\sc iii}]$<0.5$. Vermeij \& van der Hulst
(2002) estimated an electron density $n_{\rm e}\sim$ few $\times10^2$
cm$^{-3}$. At such modest electron density we would derive (Liu et al.\ 2001)
$N({\rm N})/N({\rm O})_{\rm SMC}\la0.03$. However, if the true electron
density is higher, the limit may be reduced to as low as $N({\rm N})/N({\rm
O})_{\rm SMC}<0.01$. This compares favourably with the abundance estimates
obtained by Dufour, Shields \& Talbot (1982), $N({\rm N})/N({\rm O})_{\rm
SMC}=0.036$, and suggests that, indeed, the nitrogen content in the metal-poor
SMC is even lower than that in the LMC.

%------------------------------------------------------------------------- 5.3
\subsection{Dust continuum}

As in Paper I, we define the continuum slope as follows:
\begin{equation}
\alpha \equiv 2.44\ \frac{F_\nu(85)-F_\nu(55)}{F_\nu(85)+F_\nu(55)},
\end{equation}
such that $\alpha=0$ in a flat $F_\nu$ spectrum and $\alpha=-1$ in the
Rayleigh--Jeans approximation to the long-wavelength tail of a Planck curve.
These values were computed from the mean values of the three spectral points
centered at 55 and 85-$\mu$m, respectively, and are listed in Table 4. The
error was computed by propagation of the standard deviations in the two sets
of three spectral points used.

%....................................................................... 5.3.1
\subsubsection{Dust temperatures}

The dust temperature characterizing the MIPS-SED range is a powerful
discriminant between warm circumstellar envelopes of evolved stars and cold
molecular clouds in star-forming regions. We have estimated the dust
temperature by comparison with a single grey-body:
\begin{equation}
F_\lambda=B_\lambda(T_{\rm dust})\left(1-e^{-\tau_\lambda}\right),
\end{equation}
where $B_\lambda(T_{\rm dust})$ is a Planck emission curve at the dust
temperature $T_{\rm dust}$, the optical depth $\tau_\lambda=\tau_{\rm
V}\lambda^{-\beta}$, and $\beta=1$--2, here assumed $\beta\equiv1.5$ (cf.\
Goldsmith, Bergin \& Lis 1997). Using the curve in figure 6 in Paper I, we
converted the $\alpha$ values into dust temperatures. The results are listed
in Table 4.

The temperatures of YSOs and compact H\,{\sc ii} regions in our SMC sample are
$T_{\rm dust}\approx 37$--51 K, or $T_{\rm dust}\approx 45$ K on average. This
is considerably warmer than sources in the LMC (Paper I), which have $T_{\rm
dust}\approx 32$--44 K.

On the basis of IRAS data, at lower spatial resolution than the {\it Spitzer}
data around 70 $\mu$m, Stanimirovi\'c et al.\ (2000) estimated $T_{\rm
dust}\approx 29$ K for N\,81 (which they may not have resolved), but reaching
$T_{\rm dust}\approx 37$ K in N\,88; elsewhere in the Shapley Wing they
derived $T_{\rm dust}\approx 26$--27 K, and varying between $T_{\rm
dust}\approx 28$--30 K in the SMC Bar. It thus appears that the diffuse ISM is
cold, below 30 K, but the temperature rises above 40 K within the cores of
star-forming regions. Compact sources in star-forming regions in the SMC may
have more compact, hence warmer dust envelopes than such sources in the LMC,
possibly as a result of diminished attenuation of the destructive effect of
the interstellar radiation field in the metal-poor, dust-depleted SMC.

In contrast, the dust in the two RSGs in our SMC sample is much cooler than
that in similar sources in the LMC (Paper I). In fact, the dust temperature of
the RSGs in the SMC is indistinguishable from that of the YSOs. We explore
this in more detail in the following subsection.

%....................................................................... 5.3.2
\subsubsection{Cool dust around red supergiants: swept-up ISM?}

The two bright RSGs in the LMC, WOH\,G064 and IRAS\,05280$-$6910, had MIPS-SED
spectra which declined even more steeply than a Rayleigh--Jeans tail of a hot
black-body, indicative of dust emission but at a temperature over 100 K. The
two RSGs in the SMC sample exhibit far-IR emission associated with dust of
only $T_{\rm dust}=35$--42 K. Such cool dust was occasionally seen in less
massive evolved stars in the LMC, notably carbon stars, and the explanation
put forward then was that this is swept-up interstellar dust rather than dust
produced by these stars (see Paper I).

Perhaps this scenario also applies to the two RSGs in the SMC sample (cf.\ van
Loon et al.\ 2008): if the far-IR emission from the warm circumstellar dust is
weak --- due to a low dust-to-gas ratio in the metal-poor wind --- then the
far-IR emission may be dominated by that from cool ISM dust (though the
dust-to-gas ratio in the ISM, too, is low in the metal-poor SMC). In the case
of GM\,103, the sky surrounding the point source shows complex far-IR emission
arising from the ISM near this star (see Figs.\ 1 \& 2). It would not
therefore come as a surprise that this star has collected ISM dust in a
bow-shock as it has plowed through the relatively dense ISM in this part of
the SMC. This does not appear to be the case for the other RSG, though:
BMB-B\,75 is in fact seen against a relatively ``empty'' bubble (see Figs.\ 1
\& 2).

Alternatively, the dust may have formed --- or grown (by condensation onto
pre-existing grains) --- within the bow-shock itself. Indeed, these slow
shocks (few tens of km s$^{-1}$) bear similarity to those traveling through
the molecular atmospheres of dust-producing pulsating red giant stars (Bowen
1988). This would naturally explain dust near BMB-B\,75 in the absence of
diffuse ISM dust.

%------------------------------------------------------------------------- 5.4
\subsection{Ice, molecules, and minerals}

The broad (several $\mu$m) emission features seen in some of the spectra are
almost certainly due to either minerals or ices. At the spectral resolution of
MIPS-SED, the detection of molecules is rare, though not entirely ruled out
(cf.\ Paper I and \S 5.4.2).

%....................................................................... 5.4.1
\subsubsection{Water ice}

Crystalline water ice has significant opacity in the 60--70 $\mu$m region. The
feature is generally broad, but the peak wavelength and profile shape can vary
considerably (e.g., Malfait et al.\ 1999; Dijkstra et al.\ 2006).

The bump between $\lambda\sim60$--70 $\mu$m in the spectrum of
IRAS\,00430$-$7326 is too sharp to be the peak of a single black-body, but too
broad (and displaced) to be the [O\,{\sc i}] line. We thus suggest this is a
strong emission feature of crystalline water ice. This would not be surprising
as this is also the object with the strongest water ice absorption at
$\lambda\sim3$ $\mu$m in the {\it Spitzer}-based SMC sample observed in that
way (J.M.Oliveira et al., in prep.). What is surprising is that the dust is
not very cold, but perhaps this is the reason for the ice to have crystallized
to the degree we witness.

S3MC\,00464$-$7322 is the YSO with the coldest dust in our MIPS-SED sample.
The bump around $\lambda\approx64$ $\mu$m is slightly displaced to longer
wavelengths than the [O\,{\sc i}] line seen in other objects, and it is thus
possible that in this case it is due to crystalline water ice and not oxygen.

A third YSO, S3MC\,00540$-$7321 exhibits an emission feature closer to 60
$\mu$m, which might also be due to crystalline water ice.

%....................................................................... 5.4.2
\subsubsection{Molecules}

GM\,103 displays the same sharp peak around 79 $\mu$m that was also noticed in
the MIPS-SED spectra of the SNR N\,49 and PN or high-mass star
IRAS\,05047$-$6644 in the LMC (Paper I). Then, it was suggested that it may be
due to a blend of the cluster of relatively strong transitions of water vapour
and OH emission lines around that wavelength (cf.\ Lerate et al.\ 2006). The
detection in the spectrum of GM\,103 is of rather low significance. However,
if the swept-up dust scenario is true (\S 5.3.2) then a slow (C-type) shock
could be held responsible for the prolific formation of water vapour (Bergin,
Neufeld \& Melnick 1998).

%....................................................................... 5.4.3
\subsubsection{Minerals}

Broad emission features are sometimes seen around 75 $\mu$m, most clearly in
IRAS\,01042$-$7215 and possibly in S3MC\,01051$-$7159. These are both YSOs,
just like the YSO N\,159S in the LMC in which a similar bump was seen (Paper
I).

Although the carrier of this feature is not yet known, it is reassuring that a
similar feature appears in the SMC and LMC spectra. In Paper I we suggested
that hydrous silicates provide a promising identification.

%------------------------------------------------------------------------- 5.5
\subsection{Comparison of the SMC with the LMC}

%....................................................................... 5.5.1
\subsubsection{Dust and oxygen content}

There are 11 objects in our SMC sample that are associated with star-formation
(YSOs and compact H\,{\sc ii} regions). There are 22 similar objects in the
LMC sample (Paper I). It thus becomes meaningful to compare these samples to
identify fundamental differences between these two galaxies. If differences
are found, an obvious reason for these might be the difference in metallicity
between these galaxies.

\begin{figure}
\epsscale{1.17}
\plotone{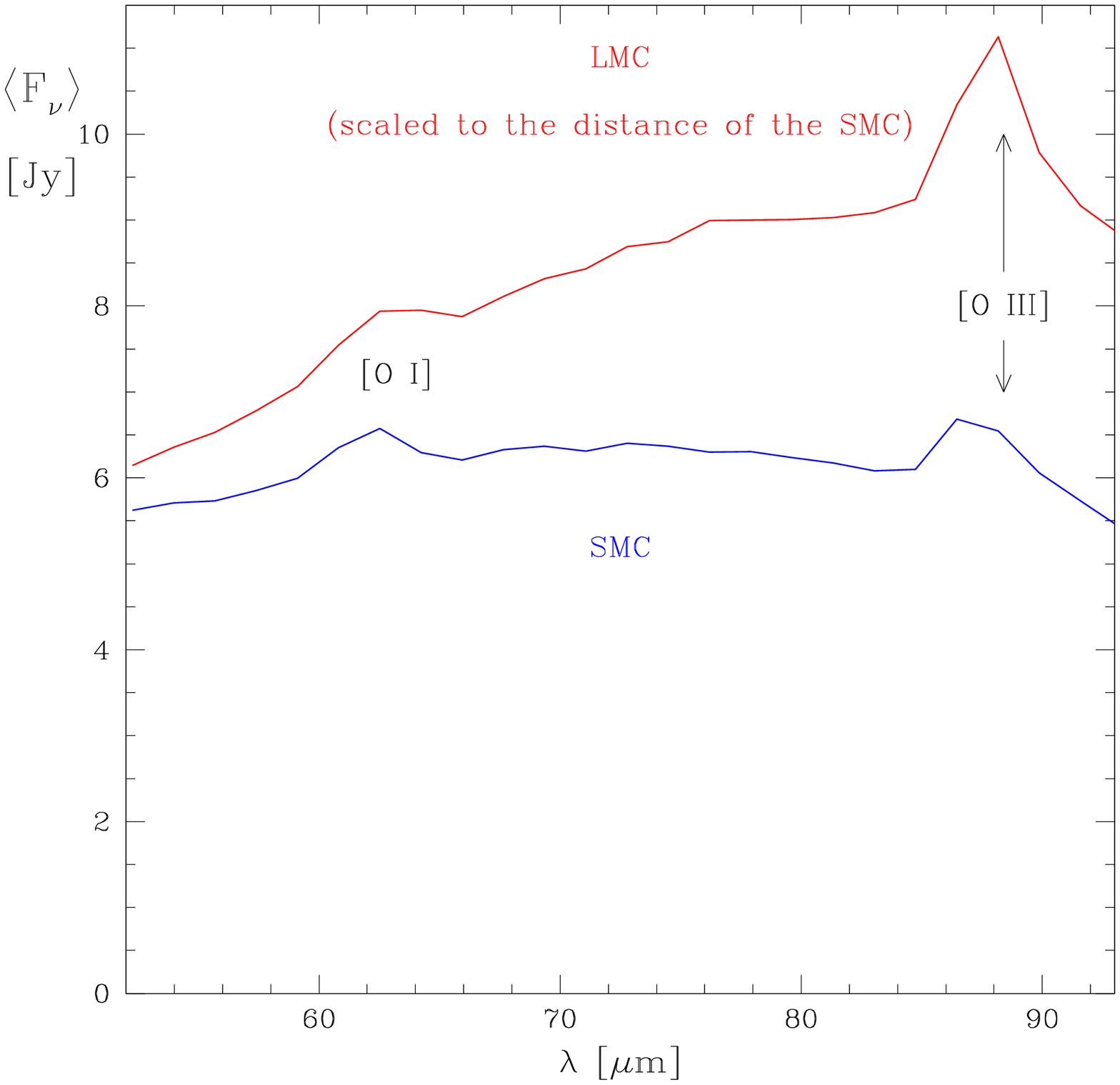}
\caption{Averaged MIPS-SED spectra of 11 objects associated with star
formation in the SMC and 22 comparable objects in the LMC. Clearly, the SMC
objects lack dust and oxygen, and their dust is warmer, compared to the more
metal-rich LMC objects.}
\label{f6}
\end{figure}

A comparison of the average MIPS-SED spectra between the SMC and LMC
star-formation objects immediately reveals three conspicuous differences
(Fig.\ 6). First, the continuum in the SMC is fainter. This could imply a
lower dust content in the SMC. Second, the continuum in the SMC is flat, as
opposed to the rising continuum in the LMC. This could imply that the dust is
warmer in the SMC, as we quantified earlier. Third, the oxygen lines are
weaker in the SMC. This could reflect a lower oxygen abundance in the SMC.

The lower dust and oxygen content is not surprising given that the metallicity
of the ISM in the SMC is lower than that of the LMC. To quantify the
difference in dust content, we compare the average far-IR luminosities between
the 11 SMC and 22 LMC star-formation objects, where we remind the reader that
we computed the far-IR luminosity due to the modified black-body emission from
dust at a temperature $T_{\rm dust}$ (see Paper I for details). This
luminosity is a function of the dust mass, $M_{\rm dust}$, as well as the dust
temperature:
\begin{equation}
L({\rm FIR}) \propto M_{\rm dust} (T_{\rm dust})^{4+\beta}.
\end{equation}
Assuming, as before, a value for $\beta\equiv1.5$, we thus obtain $\langle
M_{\rm dust}\rangle_{\rm LMC}/\langle M_{\rm dust}\rangle_{\rm
SMC}\approx3.8$.

The difference in dust content is very similar to the factor 2--4 difference
in metallicity between the SMC and LMC, and suggests that the total dust mass
of a star-forming molecular cloud core scales in proportion to its
metallicity. Whilst unsurprising at first, this seems at odds with the even
lower values for the dust-to-gas ratio that have been measured in the diffuse
ISM of the SMC ($\sim1/30$ that in the Galaxy, Stanimirovi\'c et al.\ 2000),
and with the smaller sizes of CO clouds in the SMC (Lequeux et al.\ 1994).
Grain growth within clouds might enhance the dust-to-gas ratio over that
typically encountered in the diffuse ISM; this was recently suggested to
happen in molecular cloud cores within the metal-poor tail of the SMC (Gordon
et al.\ 2009). To explain more efficient growth in the SMC would require the
densities within molecular cloud cores in the SMC to be larger than within
those in the LMC, as the growth timescale $\tau\propto N_{\rm H}^{-1}$
(Zhukovska \& Gail 2009).

For the oxygen lines we obtain a luminosity ratio of $\langle L({\rm
O})\rangle_{\rm LMC}/\langle L({\rm O})\rangle_{\rm SMC}\approx1.7$, i.e.\
very similar to the ratio of far-IR luminosities (which is $\approx1.6$) but
less than expected from a simple scaling with metallicity. Higher densities in
cloud cores in the SMC would explain this. Alternatively, if core-collapse
supernovae from massive stars contributed relatively more (compared to
intermediate-mass stars) to the chemical evolution of the SMC than they did to
that of the LMC, then the oxygen abundances of the SMC and LMC may be more
similar than their ratio of [Fe/H].

Hence we suggest that, for star formation to proceed in metal-poor clouds the
cloud cores may require higher gas densities, in order to cool efficiently and
to shield themselves sufficiently against heating by irradiation. This is in
broad agreement with theoretical models, which predict that star formation
requires higher densities at lower metallicities (Krumholz, McKee \& Tumlinson
2009).

%....................................................................... 5.5.2
\subsubsection{Dust temperature and the photo-electric effect}

\begin{figure*}
\epsscale{1.17}
\plotone{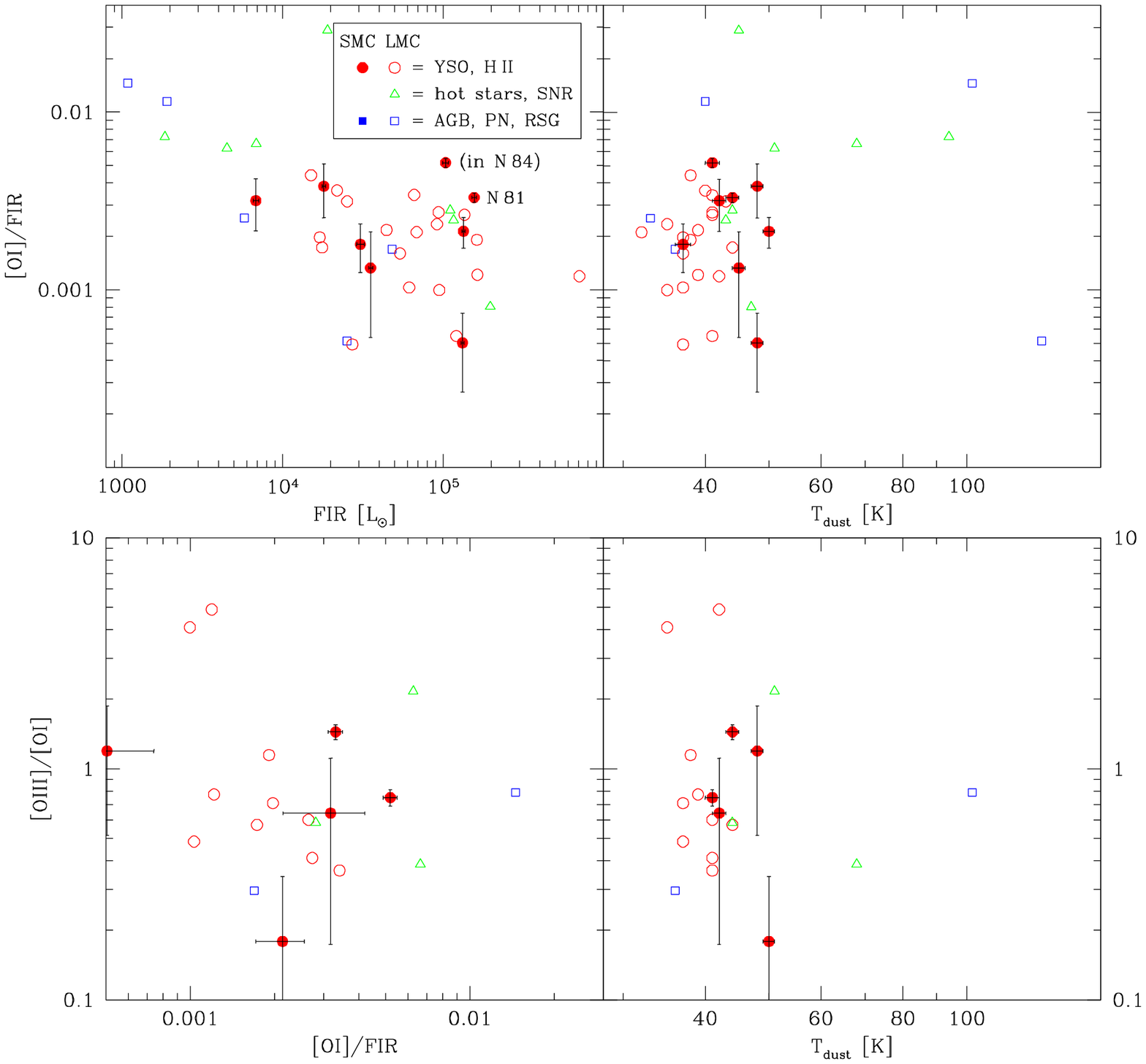}
\caption{Diagnostic diagrams utilizing the strength of the [O\,{\sc i}] and
[O\,{\sc iii}] fine-structure emission lines at $\lambda=63$ and 88 $\mu$m,
respectively, and the far-IR dust continuum (see text). The SMC data are
compared with the LMC data from Paper I.}
\label{f7}
\end{figure*}

We can also compare the SMC objects with the LMC objects in the diagnostic
diagrams introduced in Paper I (Fig.\ 7).

The difference in spectral slope is again apparent by the YSOs and compact
H\,{\sc ii} regions in the SMC exhibiting higher dust temperatures in general
than their LMC counterparts (Fig.\ 7). The reduced dust content in the SMC
causes a reduction in the ability of a molecular cloud to shield its core
against the radiation from nearby (as well as embedded) hot stars. As a
result, this radiation can penetrate deeper into the cloud, and warm it. The
lower abundance of oxygen, carbon, and molecules and dust in the SMC reduce
the ability for molecular clouds to cool. The increased heating and reduced
cooling lead to a higher equilibrium temperature of the dust. The LMC--SMC
transition in metallicity may thus probe a regime in which metallicity affects
the thermal balance of star-forming clouds (cf.\ Oliveira 2009; Oliveira et
al.\ 2009).

Although the oxygen lines are weak in the SMC stars, the [O\,{\sc i}]/FIR
ratio and [O\,{\sc iii}]/[O\,{\sc i}] line ratio are very similar between
these two galaxies (Fig.\ 7). The [O\,{\sc iii}]/[O\,{\sc i}] line ratio is a
measure of the excitation conditions in the gas. These conditions are set by a
variety of processes, distinguishing between the impact from (slow, C-type)
shocks and that from irradiation (cf.\ \S 5.1 and Paper I). On the one hand,
the boundary condition set by the interstellar radiation field (ISRF) differs
between the SMC and LMC, with the ISRF being harsher in the SMC due to the
reduced photospheric opacities in the UV of metal-poor O- and B-type stars and
the reduced extinction in the diffuse ISM as a result of the lower dust-to-gas
ratio in the metal-poor diffuse ISM (e.g., Galliano et al.\ 2005). One might
expect this to be reflected in a larger [O\,{\sc iii}]/[O\,{\sc i}] line ratio
in the SMC compared to the metal-richer LMC, which is not evident in our data.
On the other hand, even a small amount of dust can shield the interiors of
molecular clouds completely against the UV photons from the ISRF. The absorbed
heat goes into raising the dust temperature, as we indeed observe, but the
consequences for the excitation of the gas might be mild. The small sample
sizes limit our ability to explore this in more detail.

The [O\,{\sc i}]/FIR ratio is a measure of the efficiency of the
photo-electric effect. As stellar radiation impinges upon dust grains,
electrons may be released that heat the gas (Spitzer 1948). At modest
densities, $n<3\times10^3$ cm$^{-3}$, the gas is cooled mainly by radiation in
the [C\,{\sc ii}] $\lambda$158-$\mu$m line, but at higher densities cooling
through the [O\,{\sc i}] line dominates (Tielens \& Hollenbach 1985;
Weingartner \& Draine 2001). The latter case is prevalent in the compact
clouds we consider here; Rubin et al.\ (2009) found that the efficiency of the
photo-electric effect as measured through the [C\,{\sc ii}] line (in terms of
the far-IR luminosity) in the LMC decreases from $\approx1.3$\% in the diffuse
ISM down to about half that in the 30\,Doradus star-forming region. They
employed a $15^\prime$ beam for their measurements; the compact clouds that we
study here are much smaller than this, and a further reduction in the
importance of the [C\,{\sc ii}] line can be expected for these denser clouds
(cf.\ \S 5.1.2).

The efficiency of the photo-electric effect that we measured through the
[O\,{\sc i}] line (in terms of the far-IR luminosity), $\approx0.1$--0.3\%, is
not all that different between the Magellanic Clouds. This needs to be
explained, given the dependencies of the photo-electric effect on UV radiation
field ($G$), electron density ($n_{\rm e}$), gas temperature ($T_{\rm gas}$)
and grain properties (composition, size, et cetera), and the overall drop in
oxygen and dust abundance due to the reduced metallicity in the SMC compared
to that in the LMC. Weingartner \& Draine (2001) show how the efficiency
varies with grain size, for different types of grains, as a function of $\xi
\equiv G T_{\rm gas}^{1/2} n_{\rm e}^{-1}$. In a cool neutral medium, $T\ll
100$ K, a mixture of carbonaceous and silicate grains of $a\approx 0.1$ $\mu$m
achieve an efficiency of $\approx0.1$--0.3\% for $\xi\approx 10^3$ K$^{1/2}$
cm$^3$; the efficiency decreases for larger $\xi$, and increases for smaller
grains.

The similarity between the efficiency of the photo-electric effect in SMC and
LMC clouds might imply that the stronger radiation field and higher gas
temperature in the SMC (due to reduced shielding by dust) is offset by a
higher electron density (possibly the result of a higher overall gas density,
as alluded to in \S 5.5.1), to keep $\xi$ similar, or by a reduced grain size,
to offset a larger $\xi$. A higher density would be consistent with our
proposed explanation for the dependence of dust and oxygen content on
metallicity, whereas a smaller grain size would not.

%=========================================================================== 6
\section{Conclusions}

We have presented the 52--93 $\mu$m spectra of 13 compact far-IR sources in
the Small Magellanic Cloud, obtained with MIPS-SED onboard the {\it Spitzer
Space Telescope}. The sample comprises 9 Young Stellar Objects (YSOs), two
compact H{\sc ii} regions (N\,81 and a source in N\,84), and two red
supergiants (RSGs) The spectra were classified using a simple classification
scheme, introduced in Paper I. We measured the intensity of the fine-structure
lines of oxygen --- [O\,{\sc i}] at 63 $\mu$m and [O\,{\sc iii}] at 88 $\mu$m
--- as well as the slope of the dust continuum emission spectrum which we
translated into a dust temperature. The most interesting results arising from
this analysis may be summarized as follows:

\begin{itemize}
\item[$\bullet$]{We confirm the ``champagne flow'' scenario for the compact
H\,{\sc ii} region N\,81, which shows both strong [O\,{\sc i}] emission
associated with shocks and strong [O\,{\sc iii}] emission associated with
photo-ionization. We discovered a very similar source in N\,84.}
\item[$\bullet$]{The oxygen lines and dust continuum are generally weaker in
the star-forming objects in the SMC than in the LMC sample presented in Paper
I. We attribute this to the overall lower metallicity of the SMC compared to
that of the LMC. Whilst the dust mass differs in proportion to metallicity,
the oxygen mass differs less; both observations can be reconciled with higher
densities inside star-forming cloud cores in the SMC than in the LMC.}
\item[$\bullet$]{The ratio of [O\,{\sc i}] to IR (dust-processed) luminosity
is used to estimate the efficiency of photo-electric heating in the interfaces
between ionized gas and molecular clouds in YSOs and compact H\,{\sc ii}
regions; we find it is $\approx0.1$--0.3\%, i.e.\ indistinguishable from that
in the LMC (Paper I). This may be understood if the densities in cloud cores
in the SMC are indeed higher than in the LMC.}
\item[$\bullet$]{We find a very low nitrogen content of both compact H\,{\sc
ii} regions: the nitrogen-to-oxygen ratio is definitely lower than 0.04, and
perhaps even less than 0.01. This is lower than in the LMC, where it is lower
than in the Galactic Disc.}
\item[$\bullet$]{The dust in the YSOs in the SMC is warmer, 37--51 K (45 K on
average), than in the LMC (32--44 K). We propose that this is the combined
result of the diminished dust fraction at the low metallicity of the SMC,
which results in a reduced shielding against radiation, and less efficient
cooling due to the lower abundance of oxygen, molecules and dust.}
\item[$\bullet$]{The dust in both RSGs in the SMC sample is cool, in stark
contrast to the warm dust ($>100$ K) in RSGs in the LMC. We propose that this
is swept-up interstellar dust rather than dust produced in the winds of these
stars, a scenario we already proposed to explain far-IR emission from carbon
stars in the LMC (Paper I). The dust might also have grown --- or even formed
--- within the wind--ISM interaction z\^one. The presence of a bow-shock might
also explain the possible detection of water vapour and/or OH emission in the
spectrum of one of the RSGs, GM\,103. The low metallicity of the SMC is likely
to blame for the lack of detection of warmer circumstellar dust around the
RSGs.}
\item[$\bullet$]{Broad emission features are seen in a few objects; these seem
to correspond to similar features seen in the LMC sample (Paper I). They are
interpreted as due to solid state material, likely minerals and possibly
hydrous silicates.}
\item[$\bullet$]{We detect strong emission from crystalline water ice in one
YSO in our SMC sample, IRAS\,00430$-$7326, in which Oliveira et al.\ (in
prep.) already discovered very strong water ice absorption at 3 $\mu$m.
Crystalline water ice may also be visible in one or two other YSOs. Possibly,
the warmer dust causes processing of the ice.}
\end{itemize}

%==============================================================================
\acknowledgments

We thank Margaret Meixner and Ciska Kemper for their comments on Paper I,
which also helped improve this Paper II. We also thank the anonymous referee
for her/his helpful remarks. The version of the ISO data presented in this
paper correspond to the Highly Processed Data Product (HPDP) set called
'LSAN\,15700411' by Lloyd, Lerate \& Grundy (2003), available for public use
in the ISO Data Archive.

\end{document}